\documentclass[fleqn,usenatbib]{mnras}
\usepackage{newtxtext,newtxmath}
\usepackage[T1]{fontenc}
\usepackage{ae,aecompl}

\usepackage{graphicx}	% Including figure files
\usepackage{amsmath}	% Advanced maths commands
\usepackage{amssymb}	% Extra maths symbols
\newcommand{\cgg}{C_\mathrm{gg}}
\newcommand{\cgk}{C_\mathrm{g\kappa}}
\newcommand{\ckk}{C_\mathrm{\kappa\kappa}}
\newcommand{\pgg}{P_\mathrm{gg}}
\newcommand{\pgk}{P_\mathrm{g\kappa}}
\newcommand{\wgg}{w_{\mathrm{gg}}}
\newcommand{\wgk}{w_{\mathrm{g\kappa}}}
\newcommand{\qg}{q_{\mathrm{g}}}
\newcommand{\qk}{q_{\mathrm{\kappa}}}

\newcommand{\cov}{\mathbf{C}}
\title[Optimal data compression for combined probe analysis]{ Compressing combined probes: redshift weights for joint lensing and clustering analyses}

\author[Ruggeri \& Blake]{
\parbox{\textwidth}{
Rossana Ruggeri$^{1}$\thanks{Email: rruggeri@swin.edu.au}, Chris Blake$^{1}$}
\vspace*{15pt} \\
$^{1}$ Centre for Astrophysics $\&$ Supercomputing, Swinburne University of Technology, P.O. Box 218, Hawthorn, VIC 3122, Australia \\
}

\date{Accepted XXX. Received YYY; in original form ZZZ}

\pubyear{2015}

\begin{document}
\label{firstpage}
\pagerange{\pageref{firstpage}--\pageref{lastpage}}
\maketitle

% Abstract of the paper
\begin{abstract}
Combining different observational probes, such as galaxy clustering and weak lensing, is a promising technique for unveiling the physics of the Universe with upcoming dark energy experiments.
Whilst this strategy significantly improves parameter constraints, decreasing the degeneracies of individual analyses and controlling the systematics, processing data from tens of millions of galaxies is not a trivial task. 
In this work we derive and test a new estimator for joint clustering and lensing data analysis, maximising the scientific return and decreasing the computational cost. 
Our estimator compresses the data by up-weighting the components most sensitive to the parameters of interest, with no loss of information, taking into account information from the cross-correlation between the two probes.  We derive optimal redshift weights which may be applied to individual galaxies when testing a given statistic and cosmological model.
\end{abstract}

% Select between one and six entries from the list of approved keywords.
% Don't make up new ones.

%%%%%%%%%%%%%%%%%%%%%%%%%%%%%%%%%%%%%%%%%%%%%%%%%%

%%%%%%%%%%%%%%%%% BODY OF PAPER %%%%%%%%%%%%%%%%%% 

\section{Introduction}

Combining different observational probes is a promising technique to unveil the physics of the Universe with upcoming dark energy experiments. First, any tensions or inconsistencies between different probes can indicate new physics or help us correct for systematic errors not controlled in an individual analysis.  Second, a joint analysis significantly improves measurements of the parameters of interest, decreasing the degeneracies of an individual analysis  \citep{2009ApJ...695..652B, 2010A&A...523A...1J, PhysRevD.86.083504}.

The potential of these tests will be greatly enhanced by current and future cosmological surveys such as the Kilo-Degree Survey \citep{2013ExA....35...25D}, Dark Energy Survey \citep{2018ApJS..239...18A}, Hyper-Suprime-Cam (HSC) lensing survey \citep{2018PASJ...70S...4A}, Large Synoptic Survey Telescope \citep{2019ApJ...873..111I} and {\it Euclid} satellite for gravitational lensing \citep{2011arXiv1110.3193L}, and the Dark Energy Spectroscopic Instrument \citep{2019BAAS...51g..57L} and 4-metre Multi-Object Spectroscopic Telescope for galaxy clustering \citep{2012SPIE.8446E..0TD}.  Whilst this large volume of data represents a unique opportunity to understand the Universe, processing tens of millions of galaxies to detect the subtle signatures of new physics is not a trivial task.  Developing new algorithms and strategies to analyse this data is critical to maximise the outcome of these investments.
  
Further, these unprecedented data volumes create another key challenge: how do we combine information  from  galaxies  at  different  epochs  in  the  evolution  of  the  Universe? Past  analyses dealt  with  this  evolution  in  the  data  by  binning  galaxies  in  different  sub-samples  by  epoch.   However, this technique is inefficient for several reasons:  it assumes no evolution within each bin, it neglects the cross-correlation between sub-samples, and it is time-consuming because we are required to repeat the same analysis for each sub-sample of galaxies.  Moreover, systematic error may be imprinted by redshift evolution, if the same galaxy carries different weights toward different statistics in the joint analysis.

Rather than breaking the sample into multiple subsets, optimal weighting of the data is an alternative to this traditional approach which instead compresses the data, maintaining sensitivity to evolution in the sample.  Strategies for how to compress data have gained increasing attention as a powerful method to handle ``big data'',  compared to brute-force data-analysis   \citep{1997ApJ...480...22T, 2000MNRAS.317..965H}. As discussed in \citet{1997ApJ...480...22T}, optimal weighting based on the Karhunen-Lo\'eve approach can compress a data set with no loss of information, obtaining results with close-to-maximal accuracy.  In simple words, the optimal weights identify those aspects of the data that are most sensitive to the physics we care about, and amplify them with respect to other aspects of the data, which contribute mostly to the noise.  Similar to a principal component analysis,  these  weighted modes are  constructed  to  be  an  optimal  estimate  of  the  cosmological  parameters of  interest  through  the  Fisher  Information  Matrix.

\cite{1997ApJ...480...22T} discussed the need for data-compression when analysing the cosmic microwave background (CMB) with $> 10^7$ pixel  all-sky maps, where a direct numerical inversion of the covariance matrix is clearly unfeasible.
More recently, \cite{2020arXiv200506551M} discussed the application of a data-compression algorithm such as MOPED \citep{2000MNRAS.317..965H} to weak lensing measurements.
Previous studies have also developed optimal weighting schemes for data compression with focus on measuring the growth rate of structure \citep{2017MNRAS.464.2698R, 2019MNRAS.484.4100R, 2019MNRAS.483.3878R, 2019MNRAS.482.3497Z}, angular diameter distance \citep{2018MNRAS.480.1096Z},  primordial non Gaussianity \citep{2019JCAP...09..010C} and cosmic shear \citep{2019OJAp....2E..11B}.  These studies explored optimal weighting for measurements with individual probes, demonstrating  how  an  optimal weighting scheme applied to a data-set gives unbiased results and is efficient in decreasing the computational costs. 

Our current study extends the ``redshift weights'' scheme developed by \citet{2017MNRAS.464.2698R} to galaxy-galaxy lensing statistics, and the combination of lensing and clustering measurements.  When combining multiple probes, the weights for the individual probes, e.g.\ for clustering or lensing only,  ``lose their optimality'' if we neglect the cross-correlation between the different probes, which contains important information on the parameter  space  we  are  exploring.  In this work we derive and test a new weighted estimator for combining galaxy-galaxy lensing and galaxy clustering based on their covariance.  Our optimal data compression presents various advantages with respect to a more standard approach of tomographic redshift binning: by compressing the information along the redshift direction it allows for a time-efficient analysis and drastically reduces the computational time and covariance requirements, enabling us to perform data analysis over a wide redshift bin.  The weights depend on the specific cosmological statistic and fiducial model and may not be optimal for other cosmological models; however, in this case the resulting fitted parameters will remain unbiased.

The paper is organized as follows. In Sec. \ref{sec:model} we briefly describe the model for the galaxy clustering and galaxy-galaxy lensing (cross-)power spectra and covariance. In Sec. \ref{sec:weights},  we  derive the optimal weights to be applied to the lenses to optimize the statistical error of the combined probes fit. In Sec. \ref{sec:results}, we test parameter fits based on our weighted estimator using Gaussian realisations and compare the results with uncompressed analyses. In particular, we verify that the derived weights produce a lossless compression of the data and unbiased results.  Sec. \ref{sec:effbias}, we discuss a scenario in which optimal weights reduce systematic biases in fitted parameters by tracing the redshift evolution of the galaxy bias.  In Sec \ref{sec:concl}, we conclude by discussing future applications and extensions of this method.
 
\section{Models and covariance}
\label{sec:model}

\subsection{Angular power spectra for combined probes}
\label{subsec:model}

We model the angular (cross-)power spectra between two different fields $\delta_\mathrm{a}$, $\delta_\mathrm{b}$ of redshift samples $i,j$, as a function of projected Fourier mode, $\ell$, as
\begin{equation}
 C_{\mathrm{ab}}^{ij}(\ell) = \int d\chi \frac{q_\mathrm{a}^i(\chi)q_\mathrm{b}^j(\chi)}{\chi^2}P_{\mathrm{ab}}(\ell/\chi,z(\chi)),
\label{eq:cab}
\end{equation}
where $P_{\mathrm{ab}}(k,z)$ is the 3D (cross-)power spectrum of the fields at wavenumber $k$ and redshift $z$, and $\chi(z)$ is the comoving distance \citep{PhysRevD.70.043009,2017MNRAS.470.2100K}.  The weight function  $q_{a,b}(\chi)$ depends on the field considered: we focus here on auto- and cross-correlations between gravitational lensing and galaxy large-scale structure.

For the galaxy density field $\delta_\mathrm{g}$, $\qg(\chi)$ is proportional to the redshift distribution of galaxies in each bin, 
\begin{equation}
\label{eq:qg}
q_\mathrm{g}^i(\chi) =\frac{n^i_{\mathrm{lens}}(z)  }{\bar{n}_{\mathrm{lens}}^i}\frac{dz}{d\chi}\,,
\end{equation}
where $n^i_{\mathrm{lens}}(z)$ is the lens redshift distribution of sample $i$, with $z$ the redshift corresponding to $\chi$, and $\bar{n}_{\mathrm{lens}}^i$  is the average lens density.

For the convergence field $\delta_\kappa$,  $\qk(\chi)$ is  given by the lensing efficiency,
 \begin{equation}
\label{eq:qkappa}
\qk^{i}(\chi) = \frac{3 H_0^2 \Omega_m }{2 \mathrm{c}^2}\frac{\chi}{a(\chi)}\int_\chi^{\chi_{\mathrm{max}}} \, d \chi' \frac{n_{\mathrm{source}}^{i}(z)\,
%\left[ z(\chi') \right]
}{\bar{n}_{\mathrm{source}}^{i}}\frac{dz}{d\chi'} \frac{(\chi'-\chi)}{\chi'} \,,
\end{equation}
where $\Omega_m$ and $H_0$ are the values of the present-day matter density and Hubble parameter, $\chi_{\mathrm{max}}$ is the maximum comoving distance of the source distribution, and  $n^i_{\mathrm{source}}(z)$ and  $\bar{n}^i_{\mathrm{source}}$ are the source redshift distribution and average density of sources in sample $i$.  We note that Eq. \ref{eq:cab} is derived assuming the Limber and flat-sky approximations  \citep{2017JCAP...05..014L}.

\subsection{Covariance matrix }

The Gaussian covariance matrix between two angular power spectra  $ C_{\mathrm{ab}}^{ij} (\ell_1), C_{\mathrm{cd}}^{kl} (\ell_2)  $, for samples $(i,j,k,l)$ is given by \cite{PhysRevD.70.043009} and  \cite{2017MNRAS.470.2100K},
\begin{equation}\begin{split}\label{eq:covpred}
\mathbf{C}  =& \frac{4 \pi \delta_{\ell_1 \ell_2}}{ \Omega_{\rm{s}} (2\ell_1+1) \Delta \ell_1}  \times\large[ \\
&\left(C_{\mathrm{ac}}^{ik}(\ell_1)+ \delta_{ik}\delta_{\mathrm{ac}} N_{\mathrm a}^i\right) \left(C_{\mathrm{bd}}^{jl}(\ell_2)+ \delta_{jl}\delta_{\mathrm{bd}}N_{\mathrm b}^j\right) \\
+&\left(C_{\mathrm{ad}}^{il}(\ell_1)+ \delta_{il}\delta_{\mathrm{ad}} N_{\mathrm a}^i\right) \left. \left(C_{\mathrm{bc}}^{jk}(\ell_2)+ \delta_{jk}\delta_{\mathrm{bc}} N_{\mathrm b}^j\right) \right]
%\,,    
\end{split}
\end{equation}
where $\Omega_{\rm{s}}$ is the angular area of the overlapping sample in steradians.  For galaxy-galaxy lensing, the covariance of the angular power spectrum $C_{\mathrm{g\kappa}}$ depends on the $\cgg$, $\cgk$ and $\ckk$ terms.  For these probes  the noises terms are,
\begin{equation}\begin{split}
    N_{\mathrm{gg}} &= 1/\bar{n}_{\mathrm{lens}},\\
    N_{\mathrm{\kappa\kappa}} &= \sigma_e^2 /\bar{n}_{\mathrm{source}},
\end{split}    
\end{equation}
where $\sigma_e$ is the shape noise.  

\subsection{Fiducial cosmology}
\label{subsec:fidcos}

We adopt a fiducial cosmological model with matter density $\Omega_m = 0.3$, baryon density $\Omega_b=0.044$, Hubble parameter $h =0.7$, amplitude of matter clustering $\sigma_8=0.8$ and spectral index $n_s = 0.95$. 
For the galaxy bias model we choose a simple redshift-dependent relation,
\begin{equation}\label{eq:bias}
    b (z) =  b_{\mathrm{piv}}\frac{D(z_{\mathrm{piv}})}{D(z)},
\end{equation}
where $D(z)$ is the linear growth rate and we selected $b_{\mathrm{piv}} = 2$ as the value of the galaxy bias at the pivot redshift $z_{\mathrm{piv}} = 0.45$.
This relation is approximately correct for the clustering amplitude of magnitude-selected galaxy samples \citep{2001AJ....122.2267E}.  To model the galaxy-galaxy and galaxy-convergence power spectra  $\pgg$ and $\pgk$, we assume a linear bias relation where $\pgg$  $\propto b^2 \sigma_8^2$ and $\pgk \propto  b \sigma_8^2$. The power spectrum of the matter on non-linear scales is computed from CAMB \cite{Lewis:2002ah}.

\section{Optimal weights methodology}
\label{sec:weights}

We are interested in defining optimal redshift weights which average measurements from samples at different redshifts into a single final dataset containing the same information, i.e. which perform lossless data compression.  In this section we briefly introduce the optimal weights formalism and derive weights to combine galaxy clustering and galaxy-galaxy lensing measurements, $w_{gg+g\kappa}$, comparing them with individual-probe weights $w_{gg}$ and $w_{g\kappa}$.

\subsection{Derivation}
\label{sec:optweights}

\subsubsection{Optimal weights for a single parameter}

Consider a dataset ${\bf x}$ containing $n$ values, Gaussian-distributed with mean ${\bf \mu}$ and covariance $\textbf{C}$. A linear compression transforms this dataset into a single number $y$:
\begin{equation}\label{eq:lincomp}
y = \textbf{w}^T\textbf{x},
\end{equation}
where ${\bf w}$ is a vector of weights of length $n$. The compressed measurement $y$ has mean ${\bf w}^T{\bf \mu}$ and variance ${\bf w}^T \textbf{C}{\bf w}$ \citep{1997ApJ...480...22T}. 

In order to obtain lossless compression we need to select weights $\textbf{w}$ which preserve the information of the original dataset $\textbf{x}$ in the new value $y$.  More formally, such weights would conserve the Fisher information of $\textbf{x}$. Considering a single parameter of interest, e.g. $\theta_i$, we can express the Fisher information of $\theta_i$ in terms of the statistics of $y$ as,
\begin{equation} \label{eq:fishinfo}
F_{ii} = \frac{1}{2}\left(\frac{ {\bf w}^T \cov_{,i}{\bf w}}{ {\bf w}^T \cov {\bf w} } \right)^2 + \frac{\left({\bf w}^T \mu_{,i} \right)^2}{ {\bf w}^T \cov {\bf w} },
\end{equation}
where the index $,i $ denotes $\partial / \partial \theta_i $.  We note that the normalisation of the weights is arbitrary (cancels in Eq.\ \ref{eq:fishinfo}).

We select ${\bf w}$ that maximizes $F_{ii}$ in Eq. \ref{eq:fishinfo}.  A general procedure to achieve this is discussed in \citet{1997ApJ...480...22T} and \citet{2000MNRAS.317..965H}.  As is common practice, we perform our analysis for a fixed fiducial covariance matrix (e.g. evaluated from mock catalogues), independent of the model parameters, and therefore assume  $\cov_{,i}=0$ and that the information on $\theta_i$ is coming only from the second term $\propto \mu_{,i}$.  In this case, the unique solution for the weights ${\bf w}$ in Eq. \ref{eq:fishinfo} is given by, 
\begin{equation}\label{eq:weisolution}
    \mathbf{w}^T = \cov^{-1} \mu_{,i}. 
\end{equation}
Substituting Eq. \ref{eq:weisolution} into Eq. \ref{eq:lincomp}, we obtain the relation 
\begin{equation}\label{eq:yweix}
    y=  \mathbf{C}^{-1} \mu_{,i} \textbf{x}. 
\end{equation}
By substituting Eq. \ref{eq:yweix} in Eq. \ref{eq:fishinfo}, we can see that the Fisher matrix is invariant with respect to $\textbf{w}$, thus $y$ contains as much information as $\mathbf{x}$ about $\theta_i$ \citep{1997ApJ...480...22T}.

\subsubsection{Optimal weights for multiple parameters}
\label{subsec:mulparwei}

In order to determine multiple parameters from a dataset, we need to compress the dataset into multiple values to retain the information about the parameters.  We specify two equivalent approaches, following \citet{2000MNRAS.317..965H} and \citet{2019MNRAS.482.3497Z}, which lead to the same results.

Firstly, following \citet{2000MNRAS.317..965H}, we search for a second number $y'$ that contains the same information as $\textbf{x}$ about the second parameter $\theta_j$,
\begin{equation}
    y'= {\textbf{w}'}^\mathrm{T} \textbf{x}.
\end{equation}
If we require $y'$ to be uncorrelated with $y$, i.e.
\begin{equation}\label{eq:weiuncorrel}
  \textbf{w'} ^\mathrm{T}  \mathcal{\mathbf{C}} \textbf{w}^\mathrm{T} =0
\end{equation}
then, substituting Eq. \ref{eq:weiuncorrel} into Eq. \ref{eq:fishinfo}, we find the solution for $\textbf{w}'$ to be,
\begin{equation}\label{eq:weisol2}
    \textbf{w}' =\frac{ \mathbf{C}^{-1} \mathbf{\mu_{,j}} - (\mu^\mathrm{T}_{,j} \mathbf{w}) \mathbf{w} }{\sqrt{ \mu_{,j} \mathbf{C}^{-1} \mathbf{\mu_{,j} } - (\mu_{,j} \mathbf{w})^2  } }.
    \end{equation}
An alternative to this approach is described by \citet{2019MNRAS.482.3497Z}, in which a derivative matrix is defined,
\begin{equation}
     \mathbf{D} = \left(\frac{\partial \mu }{\partial \theta_i}, \frac{\partial \mu }{\partial \theta_j} \right)  
\end{equation}
and the multi-parameter weights are derived as,
 \begin{equation}\label{eq:multipwei}
     \mathbf{W} = \mathbf{C^{-1}}  \mathbf{D}, 
 \end{equation}
which generalizes Eq. \ref{eq:weisolution}.

Both these approaches provide lossless compression, leaving the Fisher matrix of the compressed sample equal to the Fisher matrix of the original data set. We compute and test the weights from both methods,   confirming that they lead to identical results.  Solutions for more than two parameters are also described by \citet{2000MNRAS.317..965H}.

\subsection{Optimal weights for $\sigma_8$}
\label{subsec:s8}

As shown above, the optimal weighting scheme depends on both the parameters of interest and the statistics used in the analysis.  As a proof-of-concept, we consider determining the single parameter $\sigma_8$ from galaxy clustering and galaxy-galaxy lensing statistics individually, and from their combination.

\subsubsection{$\cgg$ or $\cgk$ only}
\label{subsec:onestatweights}

We first consider the case of optimal weights for averaging a single statistic at given $\ell$ over redshift. For $\cgg$ the uncompressed data set $\mathbf{x}$, 
  \begin{align}
    \mathbf{x} &= \begin{pmatrix}
           \cgg(\ell, z_1) \\
           \vdots \\
           \cgg(\ell, z_n)
         \end{pmatrix}
  \end{align}
across  $n$ redshift bins, is compressed into a new data set $y$ following Eq. \ref{eq:lincomp}.  From Eq. \ref{eq:weisolution} the optimal weights for $\cgg$  have the form,
\begin{equation}\label{eq:wggindi}
 \wgg =  \cov^{-1} \partial \cgg/\partial \sigma_8,
\end{equation}
where $\cov$ is the covariance corresponding to $\textbf{x}$, i.e.\ between $ \cgg(\ell,z_i)$ and $\cgg(\ell,z_j)$, which is a diagonal matrix in the Limber approximation, and
\begin{equation}
  \frac{\partial  \cgg}{\partial \sigma_8 }= \int d\chi \frac{q_\mathrm{g}^i(\chi)q_\mathrm{g}^j(\chi)}{\chi^2} \frac{\partial P_{\mathrm{gg}}(\ell/\chi,z(\chi))}{\partial \sigma_8}.
\end{equation}
Similarly, for a data compression of the $\cgk$ power spectrum, we have
\begin{equation}\label{eq:wgkindi}
   \wgk = \cov^{-1} \partial \cgk/\partial \sigma_8,
\end{equation}
where $\cov$ is the covariance between $ \cgk(\ell,z_i)$ and $\cgk(\ell,z_j)$, and
\begin{equation}
    \frac{\partial  \cgk}{\partial \sigma_8 } = \int d\chi \frac{q_\mathrm{\kappa}^i(\chi)q_\mathrm{g}^j(\chi)}{\chi^2}\frac{\partial \pgk(\ell/\chi,z(\chi))}{\partial \sigma_8}.
\end{equation}

\subsubsection{$\cgg$ and $\cgk$ combined}
\label{subsubsec:combgkgg}

The weights determined in Sec.\ \ref{subsec:onestatweights} are optimal for individual measurements of $\cgg$ or $\cgk$ only. Since  $\cgg$ and $\cgk$ are correlated, these weights would not be optimal for data compression of the combined statistics $\cgg + \cgk$. In this section we derive the optimal weights $\textbf{w}_{gg+g\kappa}$ when compressing both $\cgg$ and $\cgk$.

We construct a data vector $\mathbf{x}$ of $2N$ measurements of $\cgg(z_i)$ and $\cgk(z_i)$, with $i = 1 \cdots N$,
  \begin{align}
    \mathbf{x} &= \begin{pmatrix}
           \cgg(\ell, z_1) \\
           \vdots \\
           \cgg(\ell, z_n)\\
           \cgk(\ell, z_1) \\
           \vdots \\
           \cgk(\ell, z_n) \\           
         \end{pmatrix},
  \end{align}
and compress this data vector into a number $y$,  
\begin{equation}\label{eq:glcompress}
    y =  \mathbf{w}_{\mathrm gg+g\kappa}^{\mathrm{T} }\mathbf{x}. 
\end{equation}
We can derive the optimal weights used in Eq. \ref{eq:glcompress} following Eq. \ref{eq:weisolution}, 
\begin{equation}
    \mathbf{w}_{gg+g\kappa} =  \mathbf{D} \cdot  \mathcal{\mathbf{C}}^{-1}
\end{equation}
with
  \begin{align}\label{eq:matrder1}
 \mathbf{D} &= \begin{pmatrix}
           \partial_{\sigma_8}\cgg(\ell, z_1)  \\
           \vdots \\
           \partial_{\sigma_8} \cgg(\ell, z_n)\\
           \partial_{\sigma_8} \cgk(\ell, z_1) \\
           \vdots \\
           \partial_{\sigma_8} \cgk(\ell, z_n) \\     
         \end{pmatrix}
  \end{align}
and
    \begin{align}\label{eq:covgggk}
 \mathcal{\mathbf{C}} &= \begin{pmatrix}
          \langle\cgg(\ell, z_1) \cgg(\ell, z_1)\rangle \dots    \langle\cgg(\ell, z_1)\cgk(\ell, z_n)\rangle\\
           \vdots \\
           \vdots \\
        \langle \cgk(\ell, z_n)\rangle  \cgg(\ell, z_1) \dots  \langle\cgk(\ell, z_n)  \cgk(\ell, z_n)\rangle \\     
         \end{pmatrix},
  \end{align}
 %   \begin{align}\label{eq:covgggk}
%  \mathcal{\mathbf{C}} &= \begin{pmatrix}
%           \langle\cgg(\ell, z_1) \cgg(\ell, z_1)\rangle \dots    \langle\cgg(\ell, z_1)\cgk(\ell, z_n)\rangle\\
%           \vdots \\
%           \vdots \\
% %
%         \langle \cgk(\ell, z_n)\rangle  \cgg(\ell, z_1) \dots  \langle\cgk(\ell, z_n)  \cgk(\ell, z_n)\rangle \\     
%          \end{pmatrix}
%   \end{align}
where these covariance matrix elements may be evaluated using Eq. \ref{eq:covpred}.
  
\subsection{Optimal weights for multiple parameters ($\sigma_8$ and $b_\mathrm{piv} $) } 
\label{subsec:s8andbias}

%\cab{(Check -- does $b$ mean $b_{piv}$ in this section?)}

Combined-probe statistics are valuable for breaking degeneracies between model parameters.  In this study we consider the proof-of-concept of using $\cgg$ and $\cgk$ to break the degeneracy between the galaxy bias $b_{\mathrm{piv}}$ and $\sigma_8$, since $\cgg \propto b_\mathrm{piv}^2 \sigma^2_8$ while $\cgk \propto b_\mathrm{piv} \sigma^2_8$.  Here we derive the optimal weighting scheme to be applied in this case, following the method described in Sec.\ \ref{subsec:mulparwei}. For simplicity we present only the derivation using the method of \citet{2019MNRAS.482.3497Z}.

We consider the data vector of measurements in Sec.\ \ref{subsubsec:combgkgg}, of length $2N$, and the covariance matrix (of dimension $2N \times 2N$) from Eq. \ref{eq:covgggk}. We generalize  Eq. \ref{eq:matrder1} to the multi-parameter case, by constructing a ($N \times 2$) matrix of the derivatives $\mathbf{D}$  of the model in each redshift bin with respect to  $\sigma_8$ and $b_\mathrm{piv}$,
  \begin{align}\label{eq:matrder2}
 \mathbf{D} &= \begin{pmatrix}
           \partial_{\sigma_8}\cgg(\ell, z_1)\;\; \partial_{b}\cgg(\ell, z_1)  \\
           \vdots \\
           \partial_{\sigma_8} \cgg(\ell, z_n) \;\;   \partial_{b} \cgg(\ell, z_n)\\
           \partial_{\sigma_8} \cgk(\ell, z_1) \;\;
           \partial_b \cgk(\ell, z_1)  \\
           \vdots \\
           \partial_{\sigma_8} \cgk(\ell, z_n) \;\;    
           \partial_{b} \cgk(\ell, z_n) \\     
         \end{pmatrix}
  \end{align}
The optimal weight matrix (of dimension $N\times 2 $) is calculated from 
\begin{equation}
    \mathbf{W}_{\mathrm  gg+g\kappa} =  \mathbf{C}^{-1}  \mathbf{D}
\end{equation}
using Eq. \ref{eq:covpred},  and the compressed data-set $\mathbf{y}$ now has dimension $2 \times 1$,
\begin{equation}
    \mathbf{y}= \mathbf{W}_{\mathrm gg+g\kappa}^{T} \mathbf{x}. 
\end{equation}

\subsection{Individual galaxy weights}

In Sec. \ref{subsec:s8} we derived weights to be applied to the power spectra measured in different redshift bins, compressing them into a single mode containing the same information as the original. These weights can be equivalently applied to individual galaxies, which can be convenient for some analyses (e.g., enabling statistics to be measured across wider redshift intervals).

Once we have determined power spectrum weights $w_{gg}$ and $w_{g\kappa}$ (or the corresponding sections of the total weight vector $w_{gg+g\kappa}$) for a particular parameter and scale, we can assign these to individual galaxies as $w_g = \sqrt{w_{gg}}$ for a clustering measurement and $w_g = w_{g\kappa}$ for a galaxy-galaxy lensing measurement.   Hence a galaxy catalogue may contain multiple weights per galaxy, where different weights are used for the measurement of different statistics. 
 This is expected as the optimal weights will always depend on the statistic under consideration. 
This recipe for applying the weights to individual galaxies has been applied in survey data analysis by e.g. \citet{2019MNRAS.483.3878R}. 
Since the weights are expected to vary slowly on the scales of interest for clustering \citep{2019MNRAS.482.3497Z},  we can choose a single effective scale instead of computing weight for every scale, which would be impractical.
 
In configuration space, if combining e.g. the shear-galaxy correlation function $\gamma_t(\theta)$ and the  galaxy-galaxy angular correlation $w(\theta)$, we can also apply weights to a pair directly, instead of to an individual galaxy \citep{2018MNRAS.480.1096Z}.

\section{Results}
\label{sec:results}

\subsection{Survey configuration}

\begin{figure}\label{subsam1}
%\begin{minipage}{2\columnwidth}
\centering
 % \begin{multicols}[3]
    \includegraphics[width=\columnwidth]{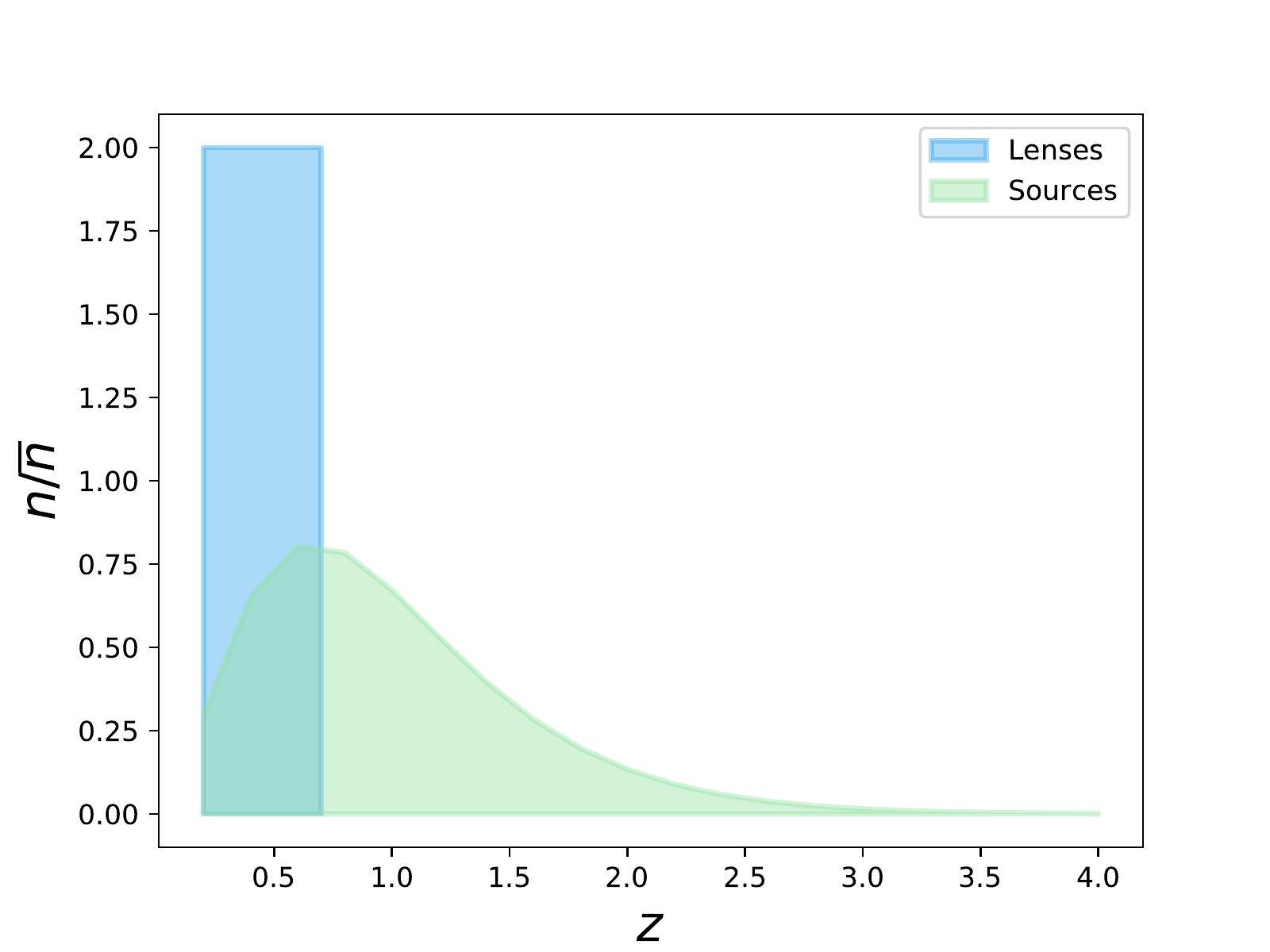}
  %\end{multicols}
%    \end{minipage}
        \caption{The source and lens redshift probability distribution of our model survey configuration.         }
    \label{fig:sourceandlensdensity}
\end{figure}

In this section we apply the data compression framework derived in Sec.\ \ref{sec:weights} to joint measurements of galaxy-galaxy lensing and galaxy clustering.  We demonstrate that an optimal weighting scheme allows for loss-less compression of the dataset, and recovers unbiased parameter constraints.

For demonstration purposes, we construct as test data a set of Gaussian realizations (see Sec.\ \ref{subsec:gausreal}) representative of current lensing and clustering surveys (we do not employ N-body simulations as we are interested in a proof-of-concept where data is precisely drawn from models).  For the lenses we assume a homogeneous galaxy sample with a constant number density distribution,
\begin{equation}
    n_{\rm lens}(z) = 10^{-4} \, h^3 {\rm Mpc}^{-3} \quad 0.2 < z < 0.7,
\end{equation}
representative of a Luminous Red Galaxy sample \citep{2001AJ....122.2267E}.  We model the redshift distribution of the sources as,
 \begin{equation}
     n_{\rm source}(z)/\overline{n}_{\rm source}(z) \propto z^2  \exp{(-z/z_0)}  \quad 0.1 < z < 3.5,
 \end{equation}
with $z_0 = 1/3$, which is representative of the Hyper-Suprime-Cam (HSC) photometric lensing catalogue \citep{2011PhRvD..83b3008O}.  These redshift probability distributions are displayed in Figure \ref{fig:sourceandlensdensity}.  The shape noise, source density and angular area used for the test data are defined in Tab.\ \ref{tab:surveyconfiguration}.  The values chosen are consistent with the HSC dataset.

\begin{table}
    \centering
    \begin{tabular}{c|c}
    \hline
        $\sigma_e$ & 0.28 \\
        $\bar{n}_{\mathrm{source}}$ & 17 arcmin$^{-2}$\\
        $\Omega_s$ & 1000 $\mathrm{deg}^2$\\
        \hline
    \end{tabular}
    \caption{ The survey configuration adopted for our test lensing dataset.  }
    \label{tab:surveyconfiguration}
\end{table}

We explored alternative survey configurations, varying the density of the lenses or their redshift range, to investigate the behaviour of the weights for different signal-to-noise ratios.  All tests performed led to equivalent conclusions, and therefore we limit our discussion to the single survey configuration described here.
  
\subsection{Gaussian realizations}
\label{subsec:gausreal}

For the test data we use a set of Gaussian realizations of the angular power spectra $\cgg$ and $\cgk$. We consider one redshift bin for the sources,  $0.0<z<3.5$, and  $N = 5$ redshift bins for the lenses of width $\Delta z = 0.1$ in the range $0.2<z<0.7$. For illustrative purposes, we select modes in the range $0 < \ell < 1000$ in bins with $\Delta \ell = 10$ ($\ell = 1000$ corresponds to $k \approx 1.7 \, h$ Mpc$^{-1}$ at $z=0.2$ and $k=0.6 \, h$ Mpc$^{-1}$ at $z=0.7$).
 
We assume measurements in different multipole bins to be independent, and for each bin $\ell$ we compute the $2N\times 2N$ covariance matrix between $\cgg(z_i, \ell)$ and $\cgk(z_j, \ell)$ (Eq. \ref{eq:covpred}), thereby including the correlations between different lens redshift slices.  To generate each Gaussian realization we Cholesky-decompose the covariance matrix $\mathbf{C}$ as 
\begin{equation}
   \mathbf{LL^*} = \mathbf{C},
\end{equation}
where $\mathbf{L}$ is a lower triangular matrix with real and positive diagonal entries,  and ``*'' denotes the conjugate transpose.  The noisy data vector $\mathbf{x}$ of each Gaussian realization for $\cgg$ and $\cgk$ is then given by 
\begin{equation}
 \mathbf{x} = \mathbf{L} \mathbf{v} + \mathbf{\mu},
\end{equation}
where $\mathbf{\mu}^\mathrm{T}= [ \cgg(z_1)\dots \cgk(z_N) ]$, and $\mathbf{v}$ is a random vector of length $2N$, drawn from a normal distribution with mean 0 and unit standard deviation.

Once the test data are created, we fit each realization for either or both of the amplitude parameters ($\sigma_8$, $b$), fixing the other cosmological parameters (the fiducial cosmology considered is listed in Sec.\ \ref{subsec:fidcos}).  We use a normal chi-squared likelihood method to perform the fit, comparing each data vector $\mathbf{x}$ with the model described in Sec.\ \ref{subsec:model}, generated in redshift slices and weighted in the same way as the data.  We quantify the errors in the fits using the standard deviation of the best-fitting parameters across 1000 different Gaussian realisations, and use these parameter errors to test the Fisher matrix predictions, the effectiveness of the loss-less data compression and the systematic errors described in Sec. \ref{sec:effbias}.

\subsection{Single parameter fit for $\sigma_8$}
\label{subsec:singleparamfit}

In this section we consider results fitting only $\sigma_8$, and fixing the other parameters to their fiducial values.  In the following section we will consider joint fits to $\sigma_8$ and $b$.  We repeat the parameter fits using three different approaches:
\begin{itemize}
\item[i)] weighted analysis,
\item[ii)] uncompressed sample analysis,
\item[iii)] wide redshift bin analysis.
\end{itemize}
Method i) uses the optimal weighting scheme to compress the information in the redshift direction into a single measurement, and method ii) corresponds to an uncompressed analysis in which the multiple redshift slices are retained and jointly analysed.   Method iii) instead utilises initial measurements in a single wide redshift bin ($0.2 < z < 0.7$ in this case), without maintaining sensitivity to the redshift evolution across the sample, which we expect to lose information.

For each method i)-iii) we consider fitting the amplitude parameters using, 
\begin{itemize}
    \item[a)] $\cgg$ only,
    \item[b)] $\cgk$ only,
    \item[c)] the combination of $\cgg$ + $\cgk$,
\end{itemize}
to investigate how the optimal weights and parameter errors depend on the statistic(s) analysed.

\paragraph*{i) Weighted analysis.}
We compress the data from each realisation by applying the optimal weighting scheme presented in Sec.\ \ref{sec:weights}.  We apply the weights to each angular power spectrum $\cgg(z_1, \ell) .. \cgk(z_n, \ell)$ as a function of $\ell$, obtaining a single mode $C_{\mathrm{gg}+ \mathrm{g\kappa}}$ 
for each $\ell$ considered.  If the compression is lossless, $C_{\mathrm{gg}+ \mathrm{g\kappa}}$ is expected to carry the same information as the uncompressed statistics  $\cgg(z_1, \ell).. \cgk(z_n, \ell)$.  We derive different weights for the different choices of statistics a), b) and c) listed above.  As discussed in Sec.\ref{sec:weights}, the optimal weights depend on the mean and the covariance matrix of the statistic(s) employed.

\paragraph*{ii) Uncompressed analysis.} We consider the angular power spectra of all $N$ redshifts slices $\cgg(z_i, \ell), \cgk(z_j, \ell)$ and the full covariance between them.  No compression of the data or optimal weighting is applied in this approach, and the $\chi^2$ function for the likelihood fitting is given by,
\begin{equation}
 \chi^2(\ell) =  \mathbf{d}^T \mathbf{C^{-1}} \mathbf{d},
\end{equation}
for each $\ell$, where $d = [ \cgg^{\mathrm{D}}(z_1,\ell) - \cgg^{\mathrm{M}}(z_1, \ell) $ $\cdots \cgk^{\mathrm{D}}(z_N, \ell) - \cgk^{\mathrm{M}}(z_N, \ell) ] $, where the superscripts D and M indicate the data and models, respectively.  The models for $\cgg$ and $\cgk$ are described in Sec.\ref{sec:model}.

\paragraph*{iii) Wide redshift bin analysis.}  Here we analyse the data considering a single wide redshift bin for the lens distribution, generating the model and data at a fixed redshift, which we take to be the mean lens redshift $z_c= 0.45$.  Otherwise, we perform fits using the same $\chi^2$ likelihood method as described above.  Comparing results from this approach with method i) demonstrates the benefit of optimal (loss-less) compression of the tomographic samples.
 
\begin{figure}
%\begin{minipage}{2\columnwidth}
\centering
\includegraphics[width=\columnwidth]{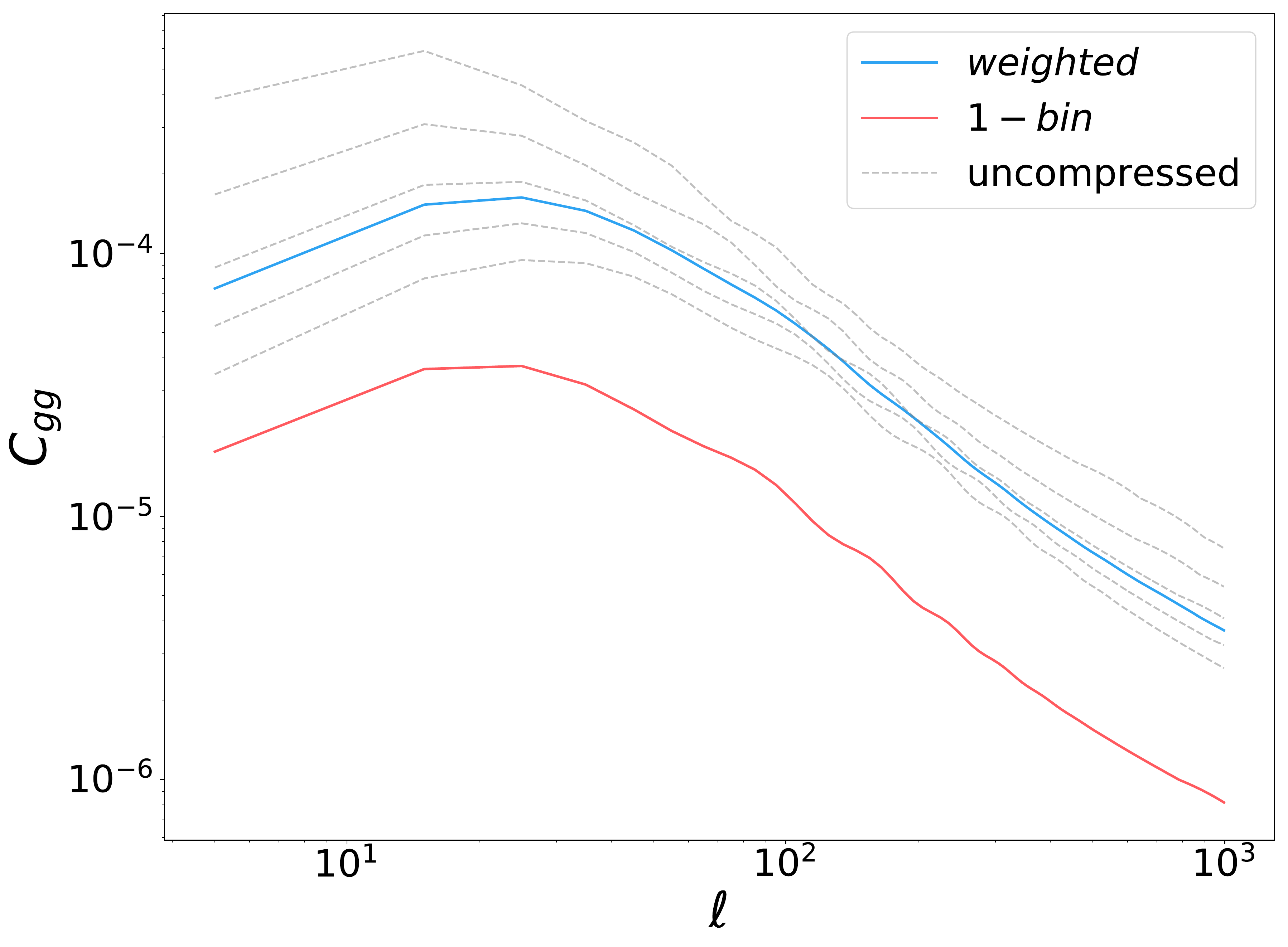} 
\includegraphics[width=\columnwidth]{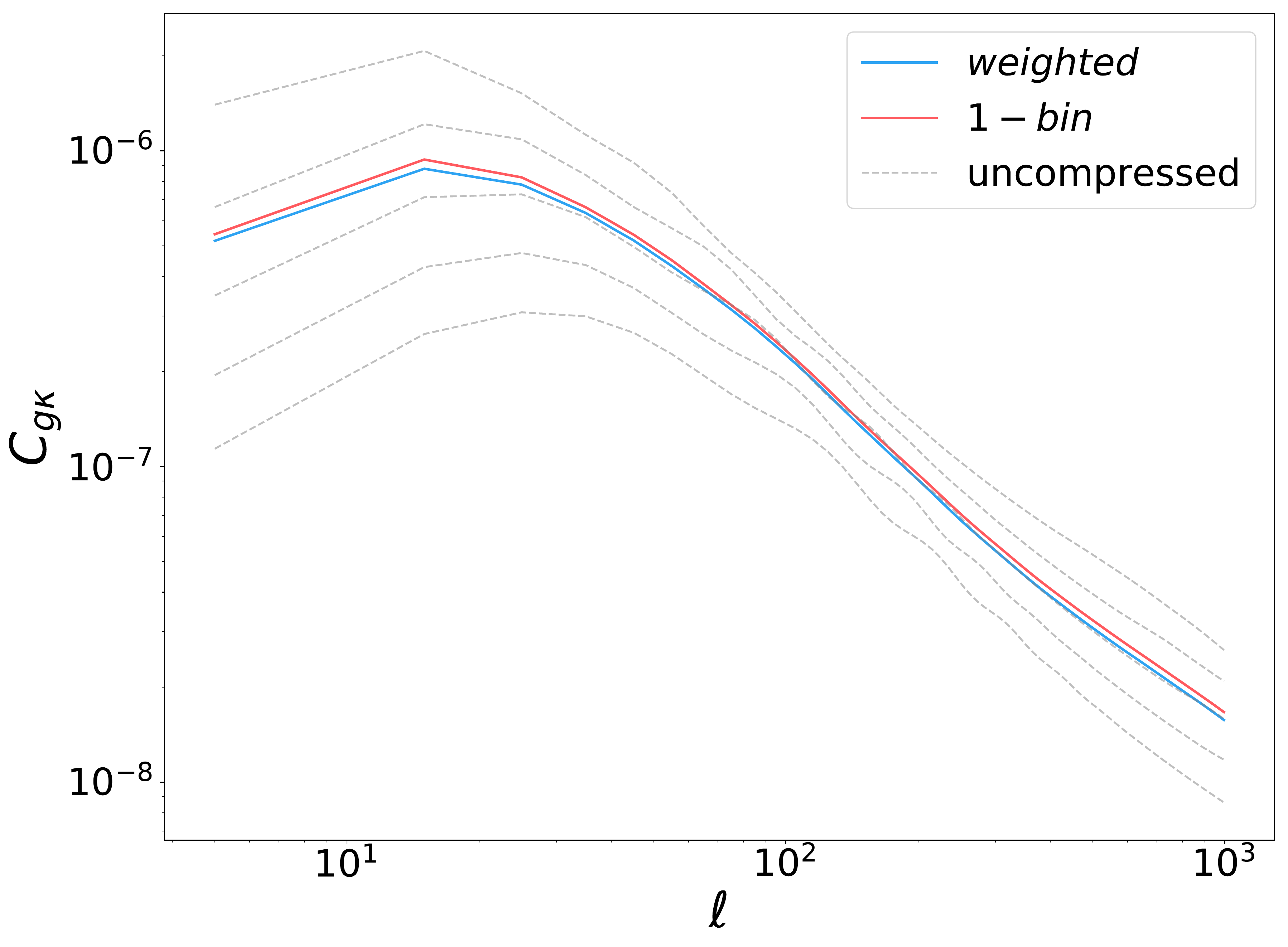} 
\includegraphics[width=\columnwidth]{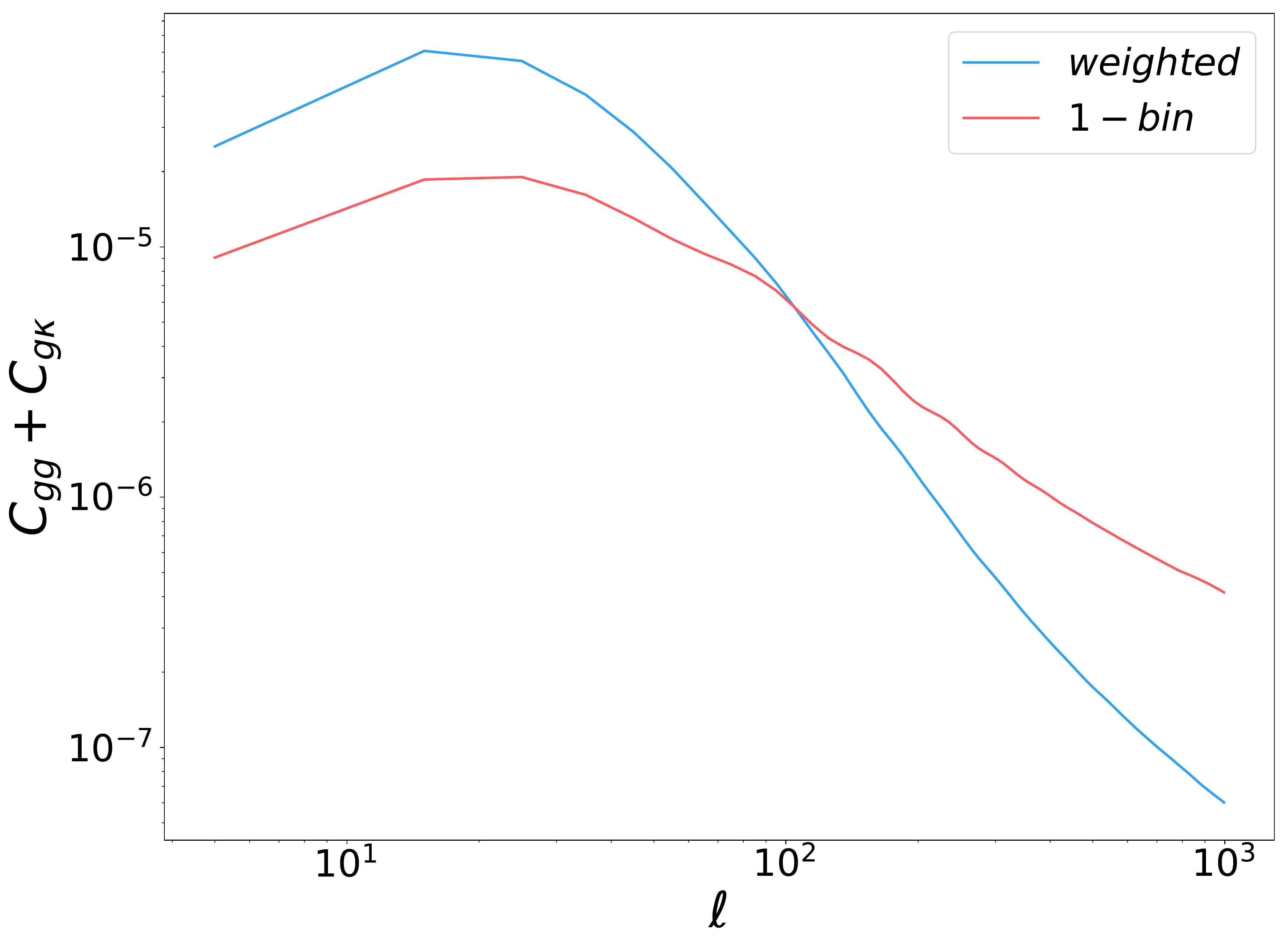} 
 % \begin{multicols}[3]
    % \includegraphics[width=.45\columnwidth]{plots/s1_cgg_cgk_ell_wnw.pdf} 
    % \includegraphics[width=.45\columnwidth]{plots/s2_cgg_cgk_ell_wnw.pdf} 
    % \includegraphics[width=.45\columnwidth]{plots/s3_cgg_cgk_ell_wnw.pdf} 
    % \includegraphics[width=.45\columnwidth]{plots/s4_cgg_cgk_ell_wnw.pdf} 
 %\end{multicols}
 %   \end{minipage}
\caption{Model galaxy clustering and galaxy-galaxy lensing angular power spectra used in this analysis.
Top panel: the galaxy clustering power spectra, $\cgg$.  The blue line indicates the weighted, compressed $\cgg$, the red line displays $\cgg$ for  one wide redshift bin, and the dotted grey lines indicate the values of $\cgg$ for each uncompressed redshift slice. 
Middle panel: the galaxy-galaxy lensing power spectra $\cgk$, with the same lines and colors as the top panel. 
Bottom panel: the summed $\cgg$ + $\cgk$ power spectra, for the weighted case (blue line) and the wide redshift bin case (red line).  
}
 \label{fig:cggcgkellwnw}

 \end{figure}

\begin{figure}
\label{fig:cggcgkweights}
%\begin{minipage}{2\columnwidth}
     \centering
    \includegraphics[width=\columnwidth]{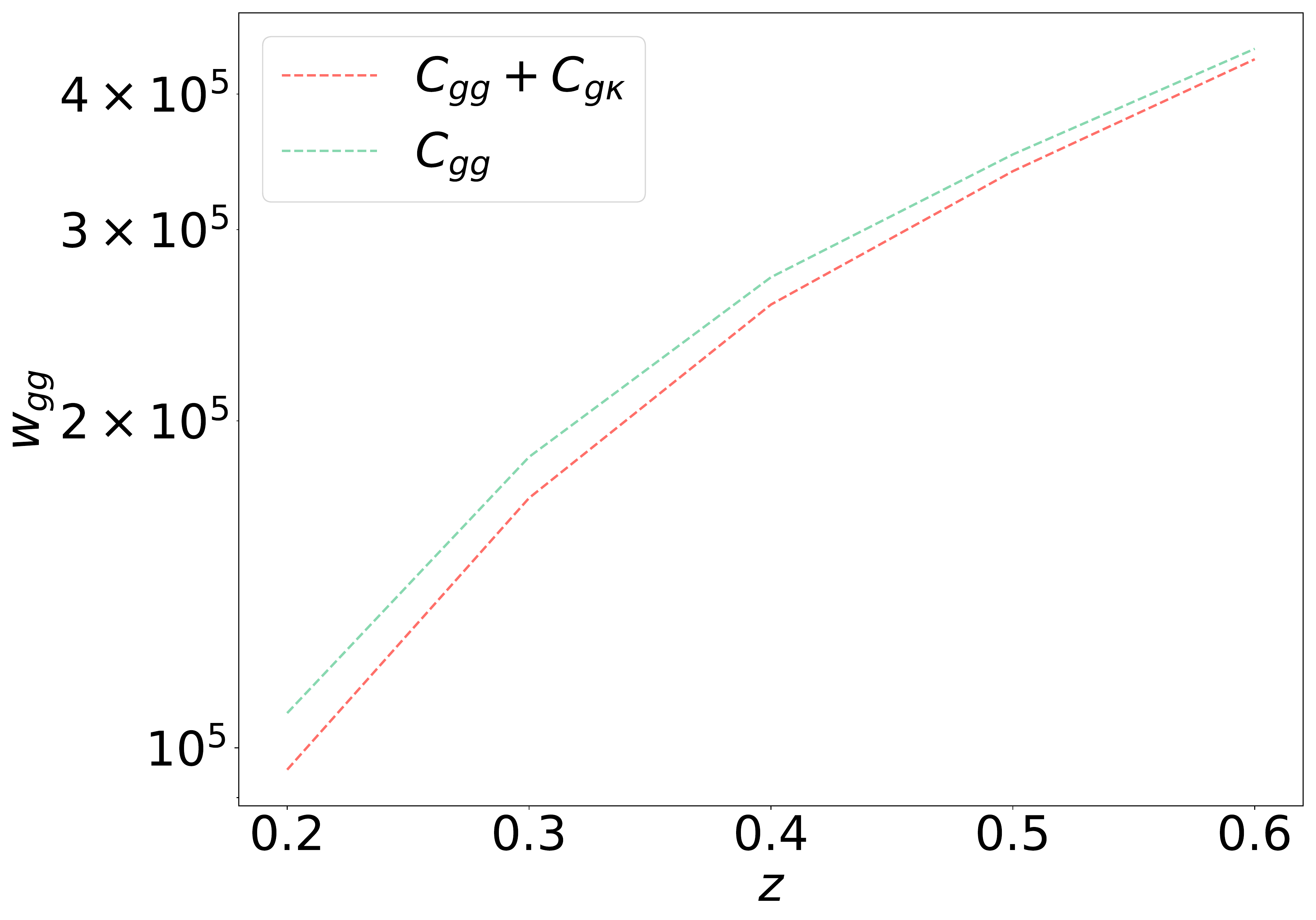}
        \includegraphics[width=\columnwidth]{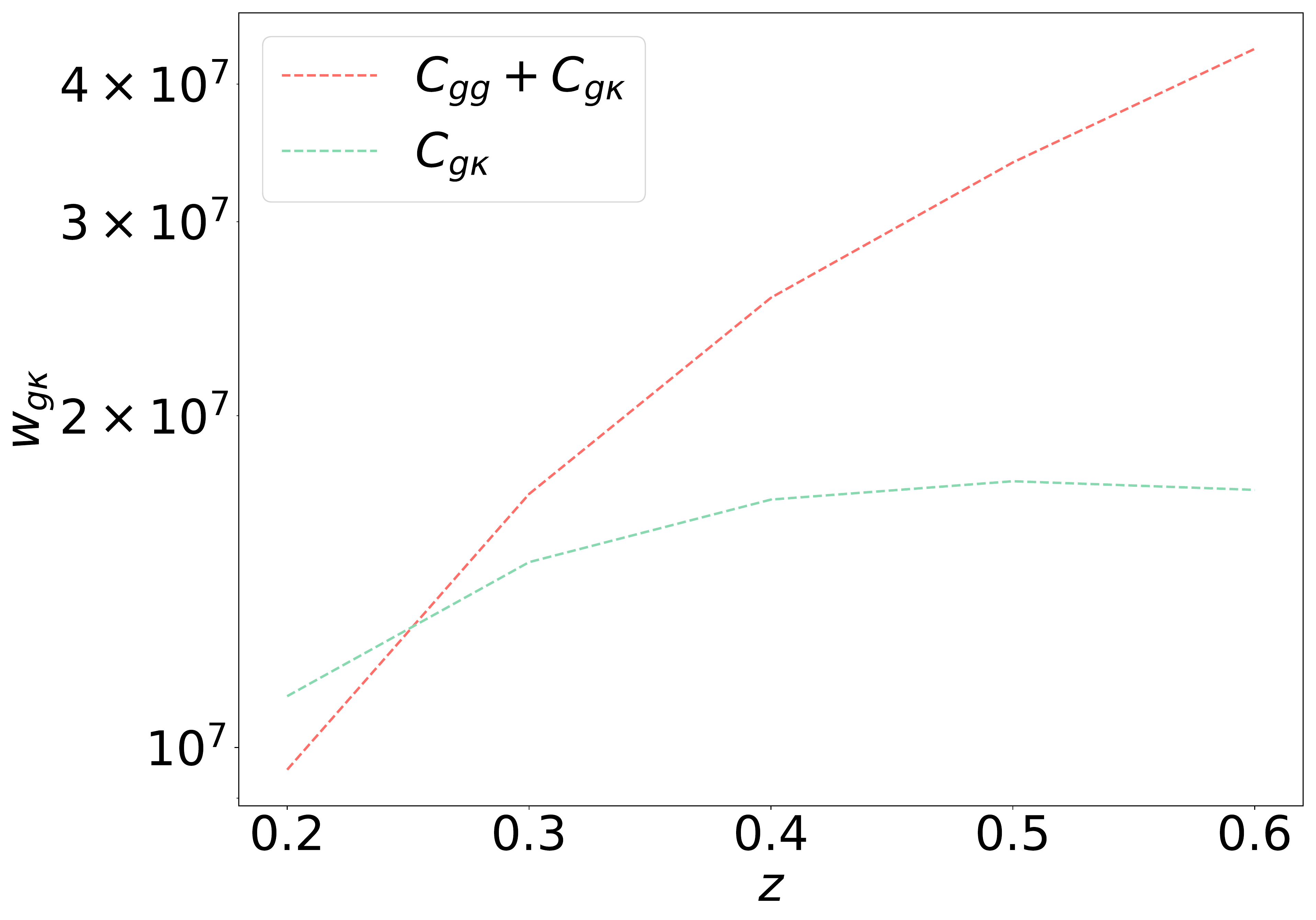}
   % \end{minipage}
\caption{The optimal weights for $\cgg$ (top panel) and $\cgk$ (bottom panel) for $\ell = 200$, as a function of lens redshift.  The red lines indicate the optimal weights when the two power spectra are combined in a joint analysis.  The green dashed lines indicate the weights for  $\cgg$ (top panel) and $\cgk$ (bottom panel) when these statistics are considered individually.  
} 
 \end{figure}

Figure \ref{fig:cggcgkellwnw} compares the angular power spectra for the analyses i) - iii) in the range $1<\ell<1000$, and Figure \ref{fig:cggcgkweights} displays the weights employed as a function of redshift when compressing statistics a), b) and c).  We note that the redshift weights applied for each statistic will be different if that statistic is analysed individually, or in combination.  We display these weights for $\ell = 200$, which corresponds to the rough location of the linear to non-linear transition at the mean redshift of the lens sample, although the weights show a similar redshift dependence for different $\ell$.  The weighting scheme does not depend on the normalization (as seen in Sec. \ref{sec:weights}), thus a convenient normalization is set for the comparison.

For the $\cgg$ statistic, the redshift weights do not vary significantly between cases a) and c).  For $\cgk$ we notice a stronger redshift dependence of the weights when moving from case b) to c).  This is due to the $\cgg$ terms in the covariance matrix in case c), enhancing the redshift sensitivity.

\begin{figure}
%\label{fig:cggcgkweights}
%\begin{minipage}{2\columnwidth} 
    \centering
    \includegraphics[width=\columnwidth]{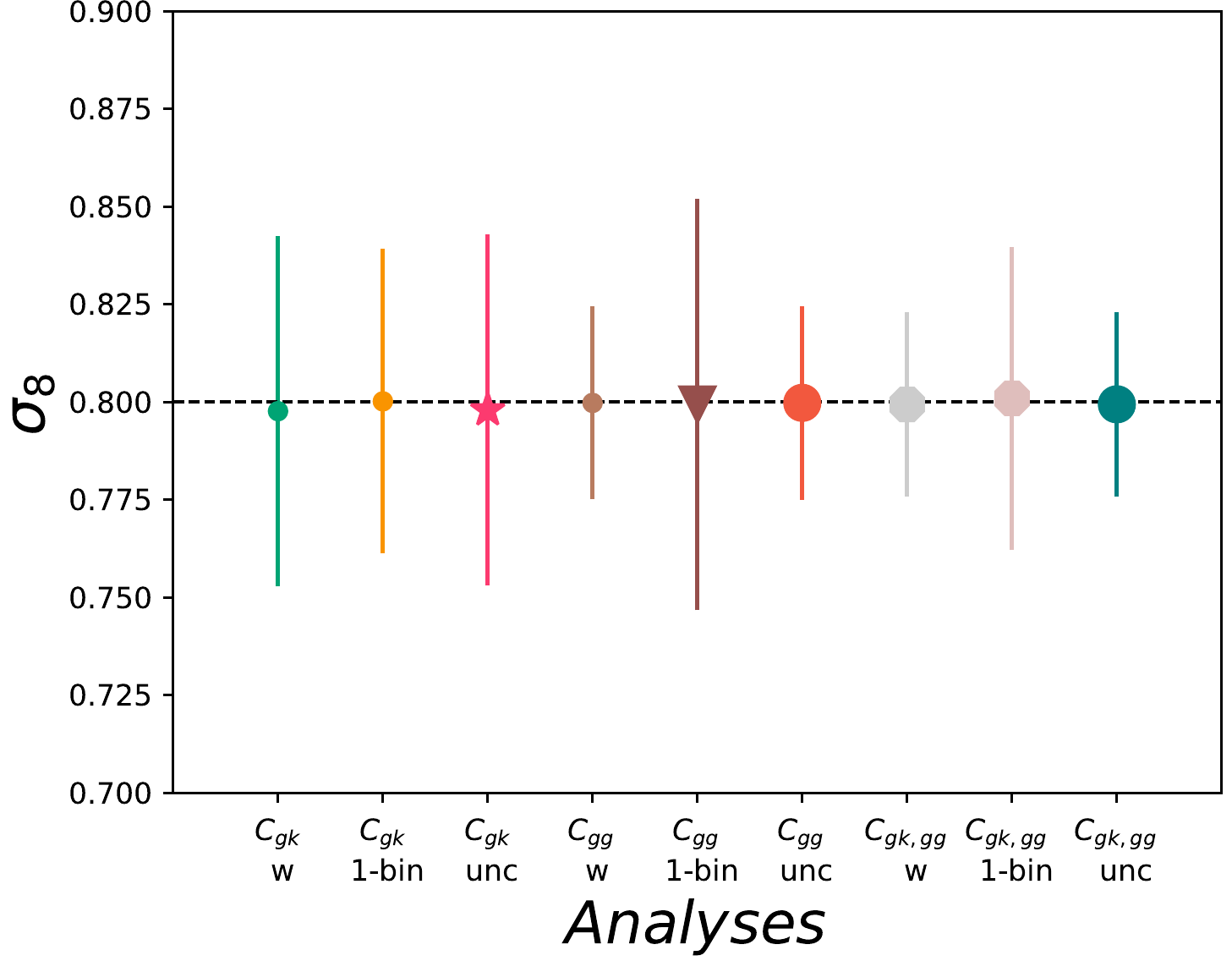} 
    % \includegraphics[width=.45\columnwidth]{plots/s2_barcha.pdf} 
    % \includegraphics[width=.45\columnwidth]{plots/s3_barcha.pdf} 
    % \includegraphics[width=.45\columnwidth]{plots/s4_barcha.pdf} 
 %   \end{minipage}
\caption{Comparison of the best-fit and standard deviation of fitting a single parameter $\sigma_8$ to the individual or jointly-analysed galaxy clustering and galaxy-galaxy lensing power spectrum, $\cgg$ and $\cgk$, for analyses of weighted and compressed data (``w''), uncompressed data (``unc''), and in a wide redshift bin (``1-bin'').}
\label{fig:fig4a}
\end{figure}

Figure \ref{fig:fig4a} presents the errors in $\sigma_8$ obtained for the three different analyses: i) weighted data, ii) uncompressed data, iii) wide-bin data, for the three different choices of statistics (cases a-c).  For each of these cases we determine the best-fit and error in $\sigma_8$ as the mean and standard deviation of the fits to each of the 1000 Gaussian realizations.  All cases considered provide an unbiased estimation of $\sigma_8$.

The analyses of the compressed datasets provide parameter errors that are comparable to those obtained in the corresponding uncompressed  analyses for all cases a-c, confirming that the compression is loss-less. 
The wide-redshift bin analysis provides the weakest constraint on $\sigma_8$, as the information from the redshift evolution is lost in this approach. This is particularly evident for $\cgg$ as its signal dilutes with the redshift bin width.

% \begin{figure}
%     \centering
%     \includegraphics[width=.9\columnwidth]{plots/s1_cgg_cgk_ell_wnw.pdf} 
%     \includegraphics[width=.9\columnwidth]{plots/s2_cgg_cgk_ell_wnw.pdf} 
%     \includegraphics[width=.9\columnwidth]{plots/s3_cgg_cgk_ell_wnw.pdf} 
%     \includegraphics[width=.9\columnwidth]{plots/s4_cgg_cgk_ell_wnw.pdf} 
%     \caption{Caption}
%     \label{fig:my_label}
% \end{figure}

\subsection{Multi-parameter fit}

We now consider jointly fitting $\sigma_8$ and the bias parameter $b_{\rm piv}$ defined in Eq.\ref{eq:bias} to our datasets.  We again compare results using the weighted, compressed datasets, the uncompressed measurements, and wide redshift bin analysis, similarly to Sec. \ref{subsec:singleparamfit}. 
 \begin{itemize}
     \item[i)] {\bf Optimal weight compression for two parameters.} 
We use the optimal weighting scheme discussed in Sec.\ \ref{subsec:s8andbias} to define weights corresponding to $\sigma_8$ and $b_{\rm piv}$ and hence compress the data and models of $\cgg(z_i), \cgk(z_j)$ into two power spectra, which we jointly analyse.
     \item[ii)]  {\bf Uncompressed analysis.} We again consider the angular power spectra for all $N$ redshift slices and their covariance when computing the $\chi^2$ statistic, adding $b_{\rm piv}$ as a free parameter in the fits.
     \item[iii)]  {\bf Wide redshift bin.}
We fit the data in one wide redshift bin, constructing our models at fixed $z= z_c$ as in Sec. \ref{subsec:singleparamfit}. We fit for $\sigma_8(z_c)$ and $b_{\rm piv}(z_c)$,  considering our model constant over redshift while the data is constructed from a model containing an evolving bias. In this way we are constructing a  test for the systematic error associated with a discrepancy between the assumed and fiducial bias evolution (see Sec. \ref{sec:effbias} for more details).

\end{itemize}

\begin{figure}
    \centering
    \includegraphics[width=\columnwidth]{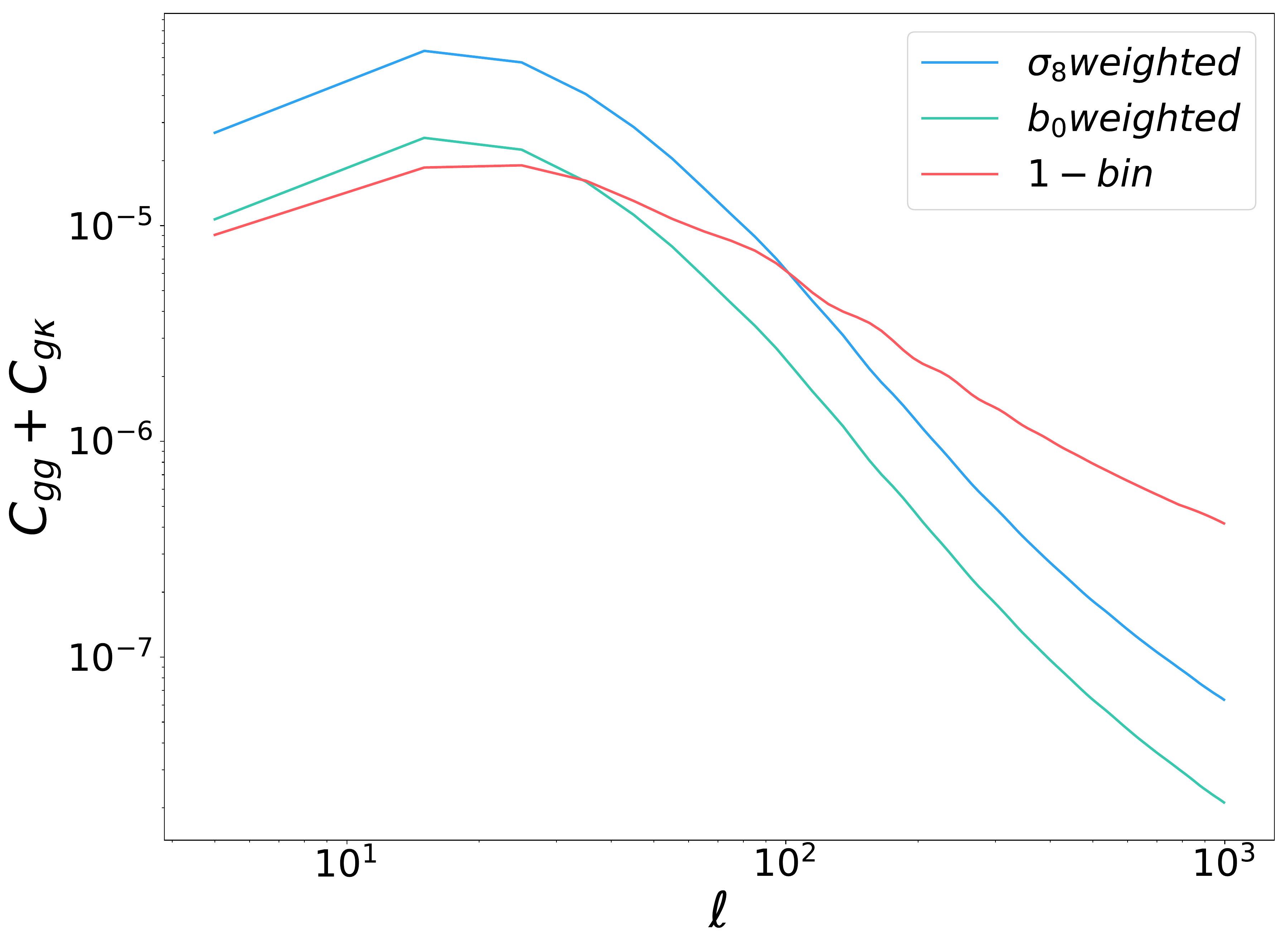}
    \caption{The sum of the $\cgg$ and $\cgk$ power spectra, comparing cases applying optimal weighting for $\sigma_8$ (blue line) and bias (green line) and a single wide redshift bin without optimal weighting (red line).}
    \label{fig:figbzpowspectra}
\end{figure}

Figure \ref{fig:figbzpowspectra} compares the  angular power spectra   for the analyses i) and iii) in the range $1<\ell<1000$. We compare the two weighted power spectra:  $w_{\sigma_8}$ (blue line) and $w_b$ (green line) with the wide-bin angular power spectrum (red line). 
The $\sigma_8$ and $b_{\rm piv}$  weighting schemes produce similar results with different amplitude.

\begin{figure}
    \centering
    \includegraphics[width=\columnwidth]{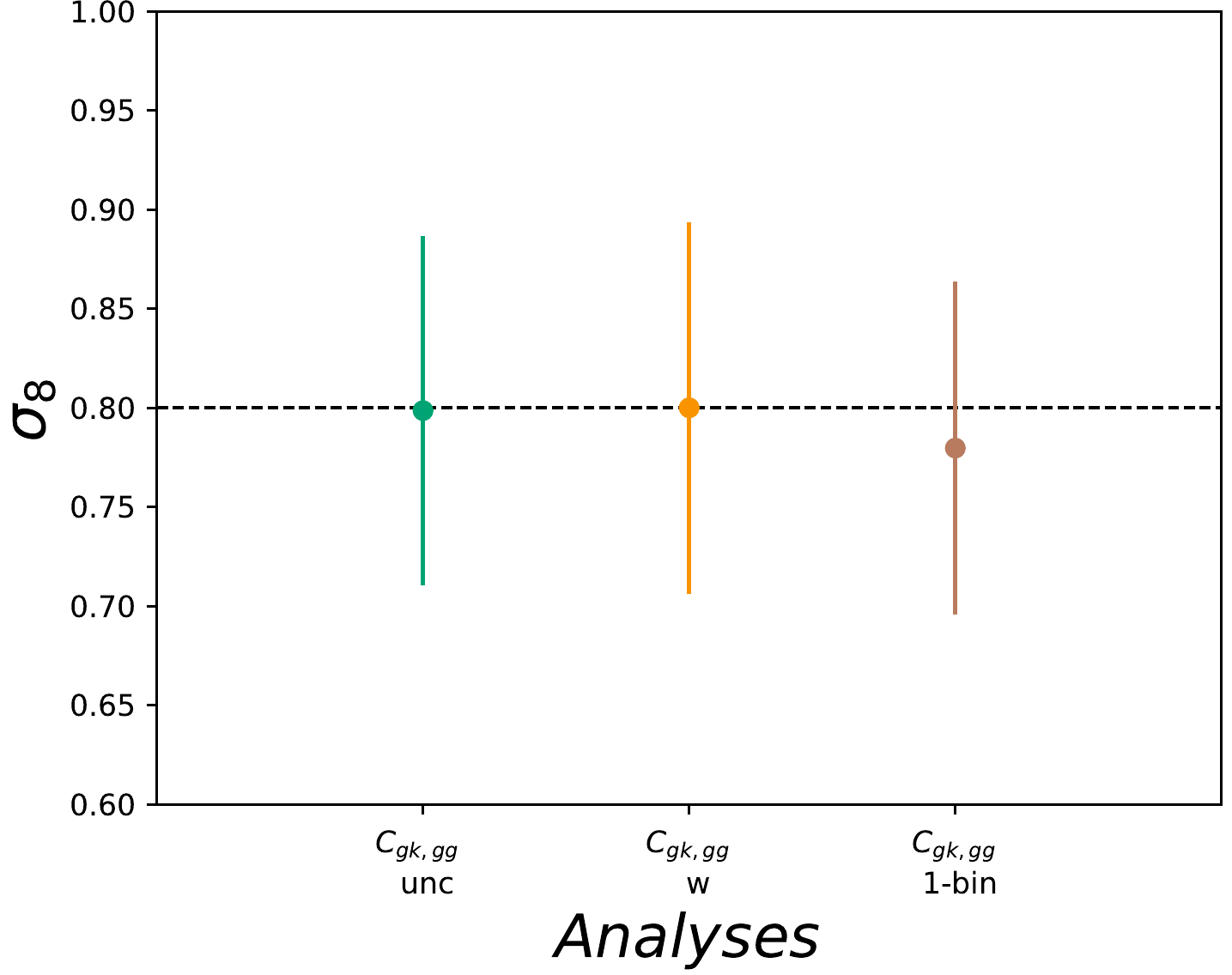} 
    % \includegraphics[width=.45\columnwidth]{plots/s2_barcha.pdf} 
    % \includegraphics[width=.45\columnwidth]{plots/s3_barcha.pdf} 
    % \includegraphics[width=.45\columnwidth]{plots/s4_barcha.pdf} 

 %   \end{minipage}
\caption{The best-fitting values and standard deviations of $\sigma_8$ for a joint fit of $\sigma_8$ and the bias parameter to the combined $\cgg$ and $\cgk$ dataset.  We compare analyses of weighted and compressed data (``w''), uncompressed data (``unc''), and in a wide redshift bin (``1-bin'').}
\label{fig:fig4c}
 \end{figure}

\begin{figure}
%\label{fig:cggcgkweights}
%\begin{minipage}{2\columnwidth} 
    \centering
    \includegraphics[width=\columnwidth]{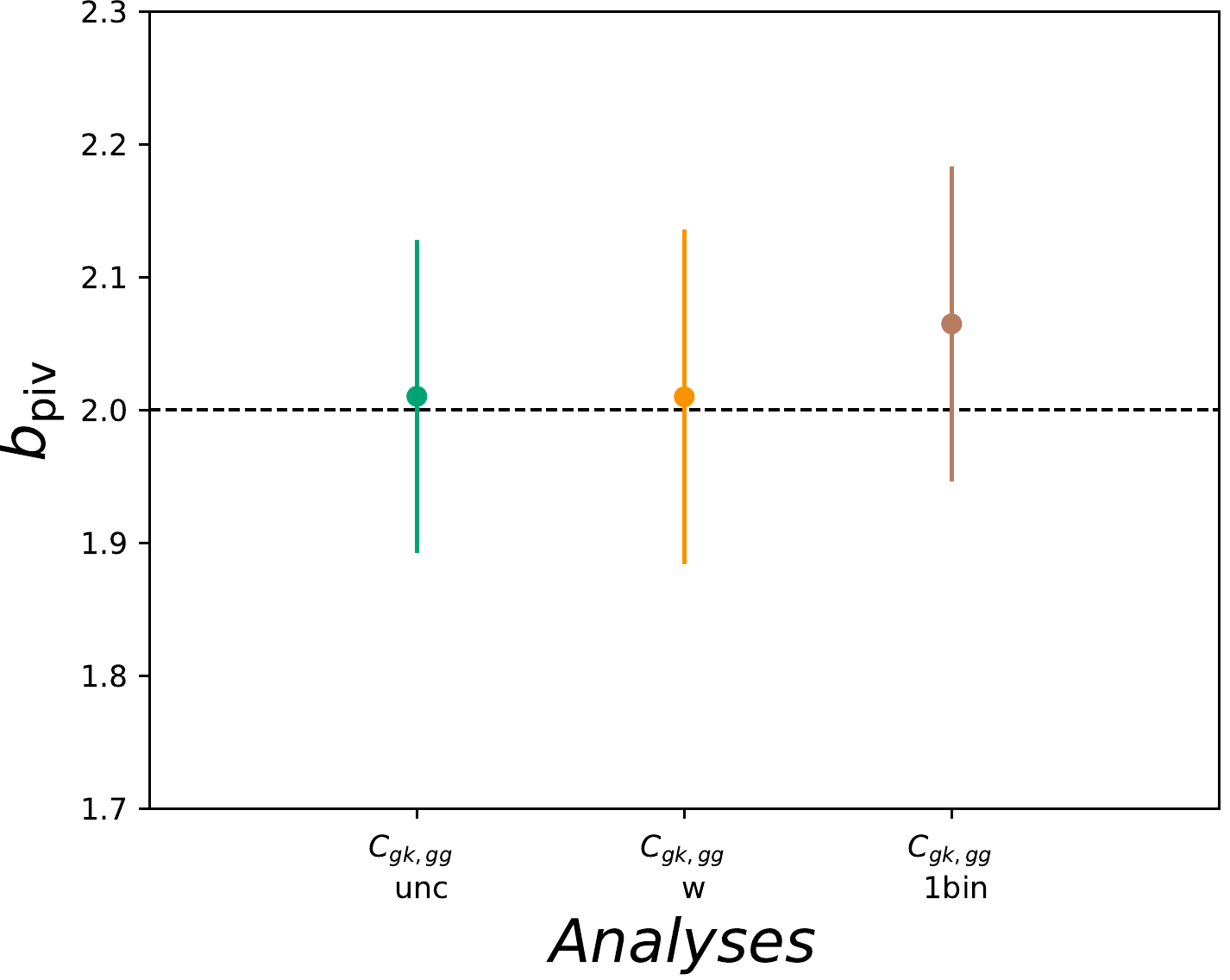} 
    % \includegraphics[width=.45\columnwidth]{plots/s2_barcha.pdf} 
    % \includegraphics[width=.45\columnwidth]{plots/s3_barcha.pdf} 
    % \includegraphics[width=.45\columnwidth]{plots/s4_barcha.pdf} 
 %   \end{minipage}
\caption{The best-fitting values and standard deviations of the bias parameter for a joint fit of $\sigma_8$ and the bias to the combined $\cgg$ and $\cgk$ dataset.  We compare analyses of weighted and compressed data (``w''), uncompressed data (``unc''), and in a wide redshift bin (``1-bin'') 
 }
\label{fig:fig4d}
\end{figure}

Figures \ref{fig:fig4c} and \ref{fig:fig4d} present the comparison between $\sigma_8$ and bias parameter fits for methods i, ii and iii, for the multi-parameter fit.  As for the single-parameter fits, the weighting and  uncompressed analysis recover unbiased estimates of $\sigma_8$.  When a wide redshift bin is used, we find a systematic error in the recovered parameters due to the discrepancy in the assumed bias evolution (see Sec. \ref{sec:effbias} for more details).


\section{Amplitude systematics}
\label{sec:effbias}

The specification of the optimal redshift weights in Sec. \ref{sec:weights} depends on the redshift evolution of the amplitudes of $\cgg$ and $\cgk$, which may not be known in advance.  Further, this redshift evolution can imprint systematic errors into parameter fits if not correctly modelled, given that it influences the signal contributed by each lens galaxy.

This issue arises due to the fact that $\cgg$ and $\cgk$ (Eq. \ref{eq:cab}) have different redshift kernels $\qg^2$ and $\qg\qk$.  Therefore, a bias factor that evolves with redshift affects the amplitude of $\cgg$ and $\cgk$ differently at the same redshift.  As a simple illustration of the effect, suppose we consider a wide redshift bin and, neglecting the redshift evolution within, generate the power spectra  in Eq. \ref{eq:cab} at a single effective redshift $z_{\mathrm{eff}}$,
\begin{equation}
\label{eq:cggint}
\cgg(\ell) = b^2 P_\mathrm{mm}(\ell/\chi_\mathrm{eff}, z_\mathrm{eff}) \int dz\;  q^2_\mathrm{g}(z),
\end{equation}
\begin{equation}
\label{eq:cgkint}
\cgk(\ell) = b P_\mathrm{mm}(\ell/\chi_\mathrm{eff}, z_\mathrm{eff}) \int dz\;  q_\mathrm{g}(z) \, q_\mathrm{k}(z).
\end{equation}
where $P_{mm}(k,z)$ is the matter power spectrum.  Here we assumed a single bias parameter for which $P_{\mathrm{gg}} = b^2 \, P_\mathrm{mm}$ and $P_{\mathrm{g\kappa}} = b \, P_\mathrm{mm}$.  If we now compare  Eq. \ref{eq:cggint} with Eq. \ref{eq:cab} we find that the galaxy bias amplitude from $\cgg$ is given by,
\begin{equation}\label{eq:bgg}
    b^2(\cgg) = \frac{\int dz  \;  q^2_\mathrm{g}(z) \, P_\mathrm{mm}(\ell/\chi , z )   }{P_\mathrm{mm}(\ell/\chi_\mathrm{eff}, z_\mathrm{eff}) \int dz  \;  q^2_\mathrm{g}(z) }. 
\end{equation}
Similarly for $\cgk$ we have,
 \begin{equation}\label{eq:bgk}
     b(\cgk)  = \frac{\int dz  \;  q_\mathrm{g}(z)  \, q_\mathrm{\kappa}(z) \, P_\mathrm{mm}(\ell/\chi , z )   }{P_\mathrm{mm}(\ell/\chi_\mathrm{eff}, z_\mathrm{eff}) \int dz  \;  q_\mathrm{g}(z) \, q_\mathrm{\kappa}(z) }.  
 \end{equation}
In general $b(\cgg) \neq b(\cgk)$, therefore fitting the galaxy bias factor from the combination of  $\cgk$ and $\cgg$ would produce  a systematic multiplicative error in the amplitude.

We quantify this systematic error by comparing fits to the uncompressed and weighted/compressed data with a wide-redshift bin analysis where the bias evolution is neglected. These results are presented in Fig. \ref{fig:fig4c} and \ref{fig:fig4d}, demonstrating how a redshift weights analysis gives unbiased results, consistent with the uncompressed sample analysis. 

To mitigate this systematic and compress the data we first need to obtain a model for the evolution of the bias. 
One approach is to divide the sample  into  narrow  redshift  bins  and fit for $b(z)$, then compress the narrow bins into a single measurement using redshift weights.  An alternative strategy, following \citet{2019MNRAS.483.3878R}, is to introduce a free functional form for the galaxy bias (e.g. a Taylor expansion).  We can then set up an iterative process, computing the first set of weights for fiducial bias parameters, fitting the parameters, and then re-generating the weights.  We note here that incorrect weights are expected to cause sub-optimality, but not bias, in the resulting parameter fits.

\section{Conclusions}
\label{sec:concl}

As cosmology transits from a data-starved science to a data-driven discipline, developing new strategies to handle the upcoming big-data volumes is a key requirement. In this work we presented a proof-of-concept of an efficient approach for combining galaxy-galaxy lensing and galaxy clustering probes across a wide redshift range in an optimal way, compressing the data-set with no loss of information. We considered just the amplitude parameters in this study, but the work could be extended to other cosmological parameters.
We derived a set of weights to constrain the galaxy bias and $\sigma_8$, to be applied to the angular power spectra $\cgg$ and $\cgk$ (and which may alternatively be applied to individual galaxies).
We test the weights on a set of Gaussian realizations mimicking the lens and source distributions of representative surveys.
We compared the weighted analysis with the uncompressed data-sets to demonstrate that the weights carry the same information as the original data-set. Finally we discussed how to handle potential systematic errors associated with evolution in redshift of the galaxy bias. 
The next step in this work is to apply the methodology to full mock catalogues and survey data samples. 

\section*{Acknowledgements}

This research was funded by the Australian Government through Australian Research Council Discovery Project DP160102705.  We thank Prof. Alexie Leauthaud and the University of California Santa Cruz for valuable discussions and hospitality during the completion of this work. We also thank Prof. Alan Heavens for  helpful suggestions.
 
%%%%%%%%%%%%%%%%%%%%%%%%%%%%%%%%%%%%%%%%%%%%%%%%%%

%%%%%%%%%%%%%%%%%%%% REFERENCES %%%%%%%%%%%%%%%%%%

% The best way to enter references is to use BibTeX:

%\bibliographystyle{mnras}
%\bibliography{example} % if your bibtex file is called example.bib

% Alternatively you could enter them by hand, like this:
% This method is tedious and prone to error if you have lots of references

%
%  These Macros are taken from the AAS TeX macro package version 4.0.
%  Include this file in your LaTeX source only if you are not using
%  the AAS TeX macro package and need to resolve the macro definitions
%  in the BibTeX entries returned by the ADS abstract service.
%
%  For more information on the AASTeX macro package, please see the URL
%	http://www.aas.org/publications/aastex.html
%  For more information about ADS abstract server, please see the URL
%	http://adswww.harvard.edu/ads_abstracts.html
%

% Abbreviations for journals.  The object here is to provide authors
% with convenient shorthands for the most "popular" (often-cited)
% journals; the author can use these markup tags without being concerned
% about the exact form of the journal abbreviation, or its formatting.
% It is up to the keeper of the macros to make sure the macros expand
% to the proper text.  If macro package writers agree to all use the
% same TeX command name, authors only have to remember one thing, and
% the style file will take care of editorial preferences.  This also
% applies when a single journal decides to revamp its abbreviating
% scheme, as happened with the ApJ (Abt 1991).

\def\jnl@style{\it}
%commente par Seb
\def\aaref@jnl#1{{\jnl@style#1}}
%ref remplace par aaref pour eviter conflit...

\def\aaref@jnl#1{{\jnl@style#1}}

\def\aj{\aaref@jnl{AJ}}                   % Astronomical Journal
\def\araa{\aaref@jnl{ARA\&A}}             % Annual Review of Astron and Astrophys
\def\apj{\aaref@jnl{ApJ}}                 % Astrophysical Journal
\def\apjl{\aaref@jnl{ApJ}}                % Astrophysical Journal, Letters
\def\apjs{\aaref@jnl{ApJS}}               % Astrophysical Journal, Supplement
\def\ao{\aaref@jnl{Appl.~Opt.}}           % Applied Optics
\def\apss{\aaref@jnl{Ap\&SS}}             % Astrophysics and Space Science
\def\aap{\aaref@jnl{A\&A}}                % Astronomy and Astrophysics
\def\aapr{\aaref@jnl{A\&A~Rev.}}          % Astronomy and Astrophysics Reviews
\def\aaps{\aaref@jnl{A\&AS}}              % Astronomy and Astrophysics, Supplement
\def\azh{\aaref@jnl{AZh}}                 % Astronomicheskii Zhurnal
\def\baas{\aaref@jnl{BAAS}}               % Bulletin of the AAS
\def\jrasc{\aaref@jnl{JRASC}}             % Journal of the RAS of Canada
\def\memras{\aaref@jnl{MmRAS}}            % Memoirs of the RAS
\def\mnras{\aaref@jnl{MNRAS}}             % Monthly Notices of the RAS
\def\pra{\aaref@jnl{Phys.~Rev.~A}}        % Physical Review A: General Physics
\def\prb{\aaref@jnl{Phys.~Rev.~B}}        % Physical Review B: Solid State
\def\prc{\aaref@jnl{Phys.~Rev.~C}}        % Physical Review C
\def\prd{\aaref@jnl{Phys.~Rev.~D}}        % Physical Review D
\def\pre{\aaref@jnl{Phys.~Rev.~E}}        % Physical Review E
\def\prl{\aaref@jnl{Phys.~Rev.~Lett.}}    % Physical Review Letters
\def\pasp{\aaref@jnl{PASP}}               % Publications of the ASP
\def\pasj{\aaref@jnl{PASJ}}               % Publications of the ASJ
\def\qjras{\aaref@jnl{QJRAS}}             % Quarterly Journal of the RAS
\def\skytel{\aaref@jnl{S\&T}}             % Sky and Telescope
\def\solphys{\aaref@jnl{Sol.~Phys.}}      % Solar Physics
\def\sovast{\aaref@jnl{Soviet~Ast.}}      % Soviet Astronomy
\def\ssr{\aaref@jnl{Space~Sci.~Rev.}}     % Space Science Reviews
\def\zap{\aaref@jnl{ZAp}}                 % Zeitschrift fuer Astrophysik
\def\nat{\aaref@jnl{Nature}}              % Nature
\def\iaucirc{\aaref@jnl{IAU~Circ.}}       % IAU Cirulars
\def\aplett{\aaref@jnl{Astrophys.~Lett.}} % Astrophysics Letters
\def\apspr{\aaref@jnl{Astrophys.~Space~Phys.~Res.}}
                % Astrophysics Space Physics Research
\def\bain{\aaref@jnl{Bull.~Astron.~Inst.~Netherlands}} 
                % Bulletin Astronomical Institute of the Netherlands
\def\fcp{\aaref@jnl{Fund.~Cosmic~Phys.}}  % Fundamental Cosmic Physics
\def\gca{\aaref@jnl{Geochim.~Cosmochim.~Acta}}   % Geochimica Cosmochimica Acta
\def\grl{\aaref@jnl{Geophys.~Res.~Lett.}} % Geophysics Research Letters
\def\jcp{\aaref@jnl{J.~Chem.~Phys.}}      % Journal of Chemical Physics
\def\jgr{\aaref@jnl{J.~Geophys.~Res.}}    % Journal of Geophysics Research
\def\jqsrt{\aaref@jnl{J.~Quant.~Spec.~Radiat.~Transf.}}
                % Journal of Quantitiative Spectroscopy and Radiative Transfer
\def\memsai{\aaref@jnl{Mem.~Soc.~Astron.~Italiana}}
                % Mem. Societa Astronomica Italiana
\def\nphysa{\aaref@jnl{Nucl.~Phys.~A}}   % Nuclear Physics A
\def\physrep{\aaref@jnl{Phys.~Rep.}}   % Physics Reports
\def\physscr{\aaref@jnl{Phys.~Scr}}   % Physica Scripta
\def\planss{\aaref@jnl{Planet.~Space~Sci.}}   % Planetary Space Science
\def\procspie{\aaref@jnl{Proc.~SPIE}}   % Proceedings of the SPIE
\def\jcap{\aaref@jnl{J. Cosmology Astropart. Phys.}}
                % Journal of Cosmology and Astroparticle Physics

\let\astap=\aap
\let\apjlett=\apjl
\let\apjsupp=\apjs
\let\applopt=\ao

\newcommand{\mpc}{\, {\rm Mpc}}
\newcommand{\kpc}{\, {\rm kpc}}
\newcommand{\hmpc}{\, h^{-1} \mpc}
\newcommand{\ihmpc}{\, h\, {\rm Mpc}^{-1}}
\newcommand{\ikms}{\, {\rm s\, km}^{-1}}
\newcommand{\kms}{\, {\rm km\, s}^{-1}}
\newcommand{\hkpc}{\, h^{-1} \kpc}
\newcommand{\lya}{Ly$\alpha$\ }
\newcommand{\lyb}{Lyman-$\beta$\ }
\newcommand{\lyaf}{Ly$\alpha$ forest}
\newcommand{\lr}{\lambda_{{\rm rest}}}
\newcommand{\bF}{\bar{F}}
\newcommand{\bS}{\bar{S}}
\newcommand{\bC}{\bar{C}}
\newcommand{\bB}{\bar{B}}
\newcommand{\vdF}{{\mathbf \delta_F}}
\newcommand{\vdS}{{\mathbf \delta_S}}
\newcommand{\vdf}{{\mathbf \delta_f}}
\newcommand{\vdn}{{\mathbf \delta_n}}
\newcommand{\vdC}{{\mathbf \delta_C}}
\newcommand{\vdX}{{\mathbf \delta_X}}
\newcommand{\xrei}{x_{rei}}
\newcommand{\lrmin}{\lambda_{{\rm rest, min}}}
\newcommand{\lrmax}{\lambda_{{\rm rest, max}}}
\newcommand{\lmin}{\lambda_{{\rm min}}}
\newcommand{\lmax}{\lambda_{{\rm max}}}
\newcommand{\hi}{\mbox{H\,{\scriptsize I}\ }}
\newcommand{\heii}{\mbox{He\,{\scriptsize II}\ }}
\newcommand{\vp}{\mathbf{p}}
\newcommand{\vq}{\mathbf{q}}
\newcommand{\vxperp}{\mathbf{x_\perp}}
\newcommand{\vkperp}{\mathbf{k_\perp}}
\newcommand{\vrperp}{\mathbf{r_\perp}}
\newcommand{\vx}{\mathbf{x}}
\newcommand{\vy}{\mathbf{y}}
\newcommand{\vk}{\mathbf{k}}
\newcommand{\vR}{\mathbf{r}}
\newcommand{\tdtwo}{\tilde{b}_{\delta^2}}
\newcommand{\tstwo}{\tilde{b}_{s^2}}
\newcommand{\tbthree}{\tilde{b}_3}
\newcommand{\tadtwo}{\tilde{a}_{\delta^2}}
\newcommand{\tastwo}{\tilde{a}_{s^2}}
\newcommand{\tabthree}{\tilde{a}_3}
\newcommand{\vnabla}{\mathbf{\nabla}}
\newcommand{\tpsi}{\tilde{\psi}}
\newcommand{\fnl}{{f_{\rm NL}}}
\newcommand{\tfnl}{{\tilde{f}_{\rm NL}}}
\newcommand{\gnl}{g_{\rm NL}}
\newcommand{\orderfour}{\mathcal{O}\left(\delta_1^4\right)}
\newcommand{\SDSSPF}{\cite{2006ApJS..163...80M}}
\newcommand{\PF}{$P_F^{\rm 1D}(k_\parallel,z)$}
\newcommand\ionalt[2]{#1$\;${\scriptsize \uppercase\expandafter{\romannumeral #2}}}%  
\newcommand{\vxone}{\mathbf{x_1}}
\newcommand{\vxtwo}{\mathbf{x_2}}
\newcommand{\vRot}{\mathbf{r_{12}}}
\newcommand{\cm}{\, {\rm cm}}

\bibliography{bibliolens}
  
%%%%%%%%%%%%%%%%%%%%%%%%%%%%%%%%%%%%%%%%%%%%%%%%%%

%%%%%%%%%%%%%%%%% APPENDICES %%%%%%%%%%%%%%%%%%%%%

% \appendix

% \section{Some extra material}

% If you want to present additional material which would interrupt the flow of the main paper,
% it can be placed in an Appendix which appears after the list of references.

%%%%%%%%%%%%%%%%%%%%%%%%%%%%%%%%%%%%%%%%%%%%%%%%%%

% Don't change these lines
\bsp	% typesetting comment
\label{lastpage}
\end{document}